# Intrinsic higher-order topological states in 2D honeycomb $\mathbb{Z}_2$ quantum spin Hall insulators

Sibin Lü and Jun Hu[*]

Institute of High Pressure Physics, School of Physical Science and Technology, Ningbo University, Ningbo 315211, China.

The exploration of topological phases remains a cutting-edge research frontier, driven by their promising potential for next-generation electronic and quantum technologies. In this work, we employ first-principles calculations and tight-binding modeling to systematically investigate the topological properties of freestanding two-dimensional (2D) honeycomb Bi, HgTe, and $Al_2O_3$(0001)-supported HgTe. Remarkably, all three systems exhibit coexistence of first-order and higher-order topological insulator states, manifested by gapless edge states in one-dimensional (1D) nanoribbons and symmetry-related corner states in zero-dimensional (0D) nanoflakes. Furthermore, fractional electron charges may accumulate at the corners of armchair-edged nanoflakes. Among these materials, HgTe/$Al_2O_3$(0001) is particularly promising due to its experimentally feasible atomic configuration and low-energy corner states. Our findings highlight the importance of exploring higher-order topological phases in $\mathbb{Z}_2$ quantum spin Hall insulators and pave the way for new possibilities in device applications.

---

[*] Email: hujun2@nbu.edu.cn

# I. Introduction

Topological insulators (TIs), characterized by the quantum spin Hall (QSH) effect, have received extraordinary attention over the past decade due to their unique electronic properties, which hold great promise for the development of new-generation electronic and spintronic devices, as well as quantum computing.[1,2,3,4] Typically, a *d*-dimensional TI manifests gapless and dissipationless boundary states in its (*d*–1)-dimensional counterpart, giving rise to a nonzero integer topological invariant number $\mathbb{Z}_2$. To date, numerous TIs have been reported in both two- and three-dimensional forms.[5,6,7] These TIs are now classified as $\mathbb{Z}_2$ or first-order TIs (FOTIs), after the recent theoretical proposition of higher-order TIs (HOTIs).[8,9,10,11] In particular, two-dimensional (2D) TIs are also known as QSH insulators. Generally, a *d*-dimensional *n*th-order TI hosts topologically nontrivial gapless boundary states in its (*d*–*n*)-dimensional form, while remaining gapped in other dimensions. For instance, 2D HOTIs display topological corner state, whereas their one-dimensional (1D) counterparts, such as nanoribbons and nanowires, do not exhibit nontrivial gapless edge states, distinguishing them from conventional 2D $\mathbb{Z}_2$ TIs.

Although HOTIs have been proposed in various theoretical models[8,9,10,11,12,13,14,15,16,17] and experimentally realized in artificially constructed acoustic and photonic systems[18], the discovery of realistic crystalline materials that exhibit HOTI states remains limited. Graphene, the first predicted 2D TI, was later predicted to be an HOTI, with topological corner states appearing in abundant geometric configurations such as triangles, rhombuses and hexagons.[19] Similarly, other carbon allotropes, including γ-graphyne and graphdiyne, that were considered as normal insulators, have also been predicted to be HOTIs.[20,21,22,23,24] Additionally, twisted bilayer graphene at specific twisting angles has been suggested to host topological corner states.[25] Furthermore, certain three-dimensional materials, such as bulk Bi and the magnetic axion insulator $Bi_{2-x}Sm_xSe_3$, have been identified as HOTIs, exhibiting hinge states in their 1D forms.[26,27] Fractal nanostructures in Bi thin films have also been observed to support robust corner states.[28] Obviously, these findings underscore the importance of further exploration of HOTIs for both fundamental research and potential applications.

Giving the abundance of conventional $\mathbb{Z}_2$ TIs, a natural question arises: can HOTI states emerge within them? In fact, bismuthene, an intrinsic 2D QSH insulator, has been predicted to exhibit topological corner state when the time reversal symmetry is broken via a magnetic substrate or an in-plane magnetic field.[29,30] In addition, in bilayer systems formed by stacking two 2D QSH insulators, interlayer coupling can induce topological corner states.[31] In these cases, the FOTI states are intentionally suppressed to facilitate the emergence of HOTI states. However, under certain circumstances, FOTI and HOTI states can transform into each other. For example, in a kagome lattice, the competition between the spin-orbit coupling (SOC) and asymmetric nearest-neighbor hopping can drive a topological phase transition between $\mathbb{Z}_2$ TIs and HOTIs.[32] Similarly, a moderate tensile strain may lead to a topological transition from an HOTI to a QSH insulator in 2D hydrogenated tetragonal stanene.[33] Intuitively, FOTI and HOTI states may coexist within $\mathbb{Z}_2$ TIs, suggesting a complex interplay between these topological phases.

In this paper, we studied the topological properties of freestanding 2D honeycomb Bi and HgTe, as well as HgTe supported on $Al_2O_3(0001)$ substrate [HgTe/$Al_2O_3(0001)$]. Our findings reveal that all three systems are intrinsic 2D $\mathbb{Z}_2$ TIs, and their zero-dimensional (0D) nanoflakes exhibit topological corner states, implying the coexistence of first-order and higher-order TI phases. In addition, the spatial distributions of the corner states are strongly influenced by the atomic configurations at the edges of the nanoflakes, potentially giving rise to fractional electron charges. Among the three systems, HgTe/$Al_2O_3(0001)$ is most promising due to its experimentally feasible geometry and the favorable energy alignment of its corner states, making it a strong contender for applications in next-generation electronic and quantum devices.

## II. Methods

The first-principles calculations were carried out in the framework of density-functional theory implemented in the Vienna ab initio simulation package (VASP).[34,35] The interaction between valence electrons and ionic cores was described using the projector augmented wave (PAW) method.[36,37] The spin-polarized local density approximate (LDA)[38] was adopted for the exchange-correlation potential and the SOC was invoked self-consistently. The 2D Brillouin zone was sampled by a 36×36 k-grid mesh. The atomic structures were fully relaxed until any component of the forces on a single atom is smaller than 0.01 eV/Å. The Bloch wavefunctions were transformed into maximally localized Wannier functions by using the WANNIER90 code.[39,40] The Wanniertools[41] and PythTB[42] packages were used to construct tight-binding models of materials considered in this work.

## III. Results and discussion

Since the 2D honeycomb lattice has been widely explored for hosting HOTI states, we focus on this type of materials. Bismuth monolayer (ML) is a prototypical 2D honeycomb lattice material with strong SOC. To explicitly elucidate the effect of SOC, we first calculated the band structure of 2D honeycomb Bi without including SOC, as shown in Fig. 1(a). It can be seen that the non-relativistic calculation predicts a direct band gap of 421 meV at the $\Gamma$ point. When the SOC is included, the band structure undergoes significant splitting, as seen in Fig. 1(b), highlighting the strong SOC effect in Bi. Moreover, the band gap transitions from direct to indirect. This is attributed to the substantial Rashba splitting of the topmost valence band near the $\Gamma$ point, which significantly raises the energy levels around the $\Gamma$ point, while the energy level at the $\Gamma$ point remains relatively unchanged. Consequently, the indirect band gap reduces to 237 meV. To investigate the topological nature of 2D honeycomb Bi, we employed the so-called "n-field" method to compute the $\mathbb{Z}_2$ topological invariant.[43,44,45] As shown in the inset of Fig. 1(b), the summation over the half Brillouin zone (shadow area) yields $\mathbb{Z}_2 = 1$, confirming that 2D honeycomb Bi is topologically nontrivial, consistent with previous report.[46]

The bands in Fig. 1 primarily originate from the Bi-6p orbitals, while the bands from the Bi-6s orbitals lie far below Fermi level ($E_F$). Accordingly, we selected the Bi-6p orbitals as the basis set to construct Wannier functions, which produced a band structure closely matching the DFT bands, as shown in Fig. 1(b). We then calculated the band structures of Bi nanoribbons. Similar to graphene, Bi nanoribbons can have either zigzag or armchair edges. Additionally, the widths of the nanoribbons must be sufficiently large to avoid coupling between edge states. Figure 2 displays the band structures of Bi nanoribbons with zigzag and armchair edges. Notably, several edge bands appear within the bulk band gap, manifesting the nontrivial topological characteristics.[46] Furthermore, the edge bands of zigzag and armchair nanoribbons exhibit distinct features due to the different arrangement of edge atoms in the two configurations.

To determine whether 2D honeycomb Bi hosts HOTI states, we investigated zero-dimensional (0D) rhombic nanoflakes, a typical geometry used to reveal HOTI corner states.[19,20] We constructed a series of 0D Bi rhombic nanoflakes with zigzag and armchair edges of varying sizes and found that well-separated corner states emerge when the edge lengths exceed approximately 50 Å. Figure 3 illustrates the energy levels of rhombic nanoflakes with zigzag and armchair edges. In the zigzag nanoflake, the energy levels of the corner states locate at around 0.23 eV above $E_F$ of 2D honeycomb Bi. We further analyzed the spatial distribution of the probability densities of their wavefunctions ($|\varphi_i|^2$). As presented in Fig. 3(a), the wavefunctions of the two corner states are separately localized at the two acute-angle corners of the nanoflake. In contrast, the armchair nanoflake exhibits distinct corner states, as depicted in Fig. 3(b). Here, four nearly degenerate energy levels are observed at approximately 0.304 eV above $E_F$ of 2D honeycomb Bi. The spatial distributions of the probability densities of these wavefunctions are almost identical, with all states localized at both acute-angle corners of the nanoflake. This behavior can be attributed to the high symmetry of the armchair nanoflake, where all edges and corners contain both types of inequivalent Bi atoms.

It is important to note that the corner states in Bi nanoflakes are positioned relatively far above $E_F$, which poses challenges for practical applications. Therefore, it is worthwhile to explore other 2D honeycomb materials that host HOTI states near $E_F$. Except for elemental materials such as graphene and Bi, several binary compounds, including BN and HgTe, also crystallize into 2D honeycomb lattice.[47] Since 2D honeycomb BN is a wide-band-gap semiconductor, we selected 2D honeycomb HgTe as a representative 2D honeycomb compound to investigate its topological properties. As shown in Fig. 4(a), the band structures of 2D honeycomb HgTe [see its atomic structure in Fig. 4(b)] without SOC exhibits graphene-like Dirac states, with bands dispersing linearly near $E_F$. Nevertheless, the Dirac states appear at the Γ point rather than the K point in graphene, featuring a small band gap of 17 meV. Analysis of their wavefunctions indicates that these Dirac states originate from hybridization between the s orbitals of Hg and Te, in contrast to graphene, where Dirac states arise from the $p_z$ orbitals of C. The parabolic band just below $E_F$ primarily derives from the in-plane $p_x$ and $p_y$ orbitals of Hg and Te. Upon including SOC, the bands undergo remarkable splitting due to the strong hybridization between the s and $p_{x/y}$ orbitals mediated by the SOC Hamiltonian. In addition, a band gap of 183 meV is induced by the SOC effect and band inversion occurs at the band gap. The calculation with "n-field" method yields $\mathbb{Z}_2 = 1$, demonstrating that 2D honeycomb HgTe is a topologically nontrivial material.

Figure 5 presents the band structures of HgTe nanoribbons with zigzag and armchair edges. Apparently, both nanoribbons exhibit edge states within the band gap of 2D honeycomb HgTe, but their characteristics differ. In the zigzag nanoribbon, four bands cross the bulk band gap, with two intersecting points at $E_F$ near the Γ point. For the armchair nanoribbon, however, there are only two edge bands with a single intersecting point located precisely at the Γ point. These edge states are obviously distinct from those in Bi nanoribbons, which can be attributed to their different atomic structures. As illustrated in the inset of Fig. 5(a), the two edges of the zigzag nanoribbon are individually terminated by Hg and Te atoms, making them structurally inequivalent. On the contrary, both edges of the armchair nanoribbon are terminated by Hg-Te dimers, as shown in the inset of Fig. 5(b), resulting in equivalent edges.

We further investigated the electronic properties of 0D HgTe rhombic nanoflakes with zigzag and armchair edges. As shown in Fig. 6, all energy levels within the energy range are doubly degenerate due to the reduced structural symmetry compared to Bi nanoflakes. Interestingly, each structure hosts two corner states, with their energy levels positioned just above $E_F$, as highlighted by the red dots in Fig. 6. They are the energy levels of the lowest unoccupied molecular orbitals (LUMOs), separated from the highest occupied molecular orbitals (HOMOs) by 41 and 114 meV, respectively for the zigzag and armchair nanoflakes. Obviously, these energy levels are much lower than those in Bi nanoflakes, so that they are much easier to be manipulated for applications. The spatial distributions of the two corner states in each structure are identical. In the zigzag nanoflake, both corner states are localized at the acute-angle corner which is terminated by a Te atom, as seen in the set of Fig. 6(a). Moreover, it can be seen that the Hg atoms at the corner region contribute most of the corner-state wavefunctions, while only the edge Te atoms near the corner exhibit notable contributions. In contrast, the corner states in the armchair nanoflake primarily originate from the edge Te atoms near the two blunt-angle corners. It is worth noting that the spatial distribution of the corner states in the armchair nanoflake appears less localized than those in the zigzag nanoflake. This is likely due to the limited nanoflake size in our calculations, as computational demands increase rapidly with system size. Nevertheless, the primary localization characteristics of the corner states are clearly demonstrated.

Given that freestanding 2D TIs including HgTe are challenging to obtain, it is crucial to identify suitable supporting substrates for potential applications in realistic electronic devices. It has been revealed that $Al_2O_3$(0001) is an ideal substrate for stabilizing 2D honeycomb HgTe.[47] As illustrated in Fig. 7(a), HgTe/$Al_2O_3$(0001) prefers a configuration in which each Hg and Te atom is bonded to three oxygen atoms and one aluminum atom, respectively. The short bond lengths of Hg-O and Te-Al bonds, i.e. $d_1$

= 2.70 Å and $d_2$ = 2.74 Å, indicate strong interaction between HgTe layer and Al$_2$O$_3$(0001) substrate. Consequently, the system should no longer be treated as separate entities but rather as an integrated whole. Interestingly, the key feature of the band structure near the Γ point remain intact, as shown in Fig. 7(b) and 7(c), because Al$_2$O$_3$(0001) has a wide band gap with its valence band maximum at approximately -0.8 eV below $E_F$ and its conduction band minimum at about 4.2 eV above $E_F$. In this system, the SOC-induced band gap is 239 meV, and its topological characteristic is preserved with $\mathbb{Z}_2 = 1$. Accordingly, it is appropriate to selecte the orbitals of HgTe for further investigating the electronic properties of Al$_2$O$_3$(0001)-supported HgTe nanoribbons and nanoflakes. The bands obtained from Wannier function fitting accurately reproduce the corresponding DFT bands, as plotted in Fig. 7(c), confirming that the chosen basis set is indeed sufficient.

The band structures of Al$_2$O$_3$(0001)-supported HgTe nanoribbons with zigzag and armchair edges are depicted in Fig. 8. Although they differ from those of freestanding HgTe nanoribbons, the main characteristics of the edge bands are preserved, manifesting that HgTe/Al$_2$O$_3$(0001) indeed retains its TI properties. As displayed in Fig. 9, the 0D rhombic nanoflakes also exhibit double-degenerate energy levels, similar to their freestanding counterparts. However, the corner states in both cases are predominantly contributed by Te atoms, while Hg atoms have ignorable contribution. Specifically, the corner states in the zigzag nanoflake arise from Te atoms located near one of the two acute-angle corners, whereas those in armchair nanoflake originate from Te atoms positioned along the edges surrounding the two blunt-angle corners symmetrically. In addition, the corner-state energy levels are 19 and 47 meV above $E_F$, respectively for the zigzag and armchair nanoflakes, further reduced compared to those in freestanding HgTe nanoflakes. Such energy levels facilitate low excitation energies for transport in potential device applications.

It is known that integrating the probability density of a wavefunction yields the number of electrons associated with that wavefunction, which should be exactly equal to 1 according to quantum mechanics. In the case of armchair nanoflakes of the materials considered in this work, the probability density of an individual corner-state wavefunction is distributed across two opposite corners. However, the two parts are completely separated by the gapped interior region, as shown in Fig. 3(b), 6(b) and 9(c). Consequently, each related corner carries half electron charge, i.e. fractional electron charge of $\frac{1}{2}e$ might be observed in armchair nanoflakes. In fact, similar fractional electron charges have been reported in graphene nanoflakes, where electron charges of $\frac{1}{2}e$ and $\frac{1}{3}e$ emerge depending on the specific geometries of nanoflakes.[19] It is important to emphasize that, despite the spatial separation, the wavefunction of a single corner state remains an indivisible quantum entity, with the two fractional electron charges intrinsically entangled. Nevertheless, for certain applications requiring localized charge manipulation, it may be desirable to concentrate the entire electron charge at a single corner. This can be achieved through the application of local perturbations, such as site-specific magnetization or electric potential, to break the symmetry connected by the two corners, as proposed in recent studies.[22] These findings underscore the interplay between geometric confinement, quantum entanglement, and external tunability in the design of functional higher-order topological systems.

## Ⅳ. Conclusions

In summary, we have systematically investigated the topological properties of 2D honeycomb Bi, HgTe and HgTe/Al$_2$O$_3$(0001), based on first-principles calculations and tight-binding simulations. All three systems are intrinsic 2D $\mathbb{Z}_2$ TIs, manifested by the edge states in 1D nanoribbons. Additionally, they exhibit corner states in 0D rhombic nanoflakes, implying the coexistence of first-order and higher-order TI states. Notably, the corner states in zigzag nanoflakes are localized at a single corner, whereas those in armchair nanoflakes are distributed across two counter corners. This spatial separation in armchair configurations enables the observation of fractional electron charge of $\frac{1}{2}e$, , a phenomenon supported by the bipartite nature of the wavefunctions. However, freestanding 2D honeycomb Bi and HgTe present significant limitations: the corner states in Bi are energetically distant from $E_F$, and the synthesis of freestanding HgTe remains experimentally challenging. In contrast, the HgTe/Al$_2$O$_3$(0001) system offers a promising alternative, as its atomic configuration is experimentally feasible, and its corner states are energetically aligned close to $E_F$. These attributes make HgTe/Al$_2$O$_3$(0001) a promising candidate for next-generation electronic and quantum devices.

**Acknowledgements**
This work is supported by the Program for Science and Technology Innovation Team in Zhejiang (Grant No. 2021R01004), the Six Talent Peaks Project of Jiangsu Province

(2019-XCL-081), the start-up funding of Ningbo University and Yongjiang Recruitment Project (432200942).

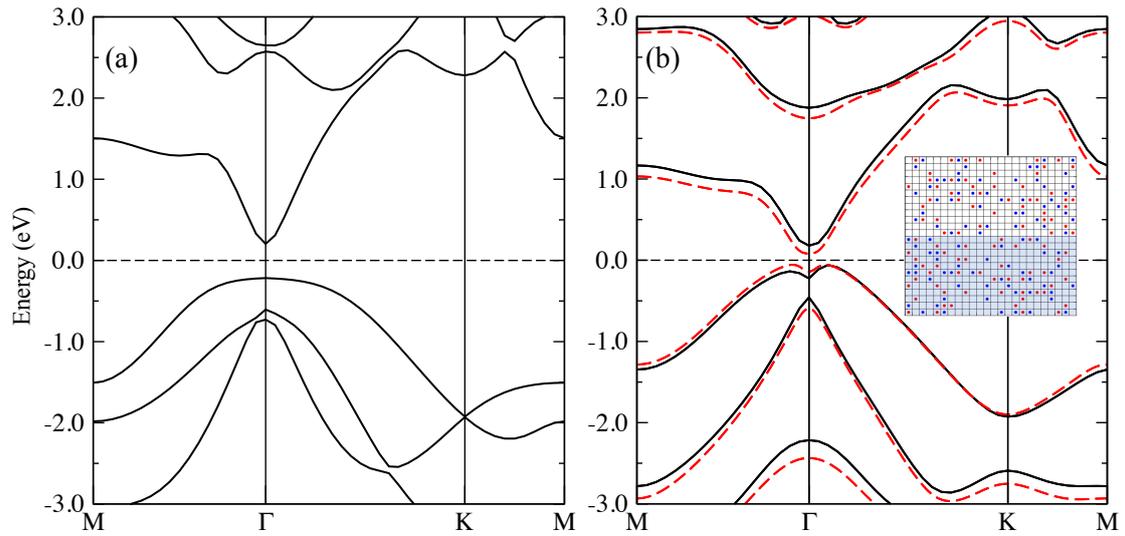

Fig. 1. Band structures of 2D honeycomb Bi (a) without and (b) with spin-orbit coupling. The red dashed curves in (b) are the energy bands from tight-binding model with the fitted basis set of Wannier functions. The Fermi level (dashed horizontal line) is set as the zero energy. The inset in (b) is the $n$-field configuration in the Brillouin zone sampled by a $24 \times 24$ k-grid mesh. The nonzero points are denoted by red ($n = 1$) and blue ($n = -1$) dots, respectively.

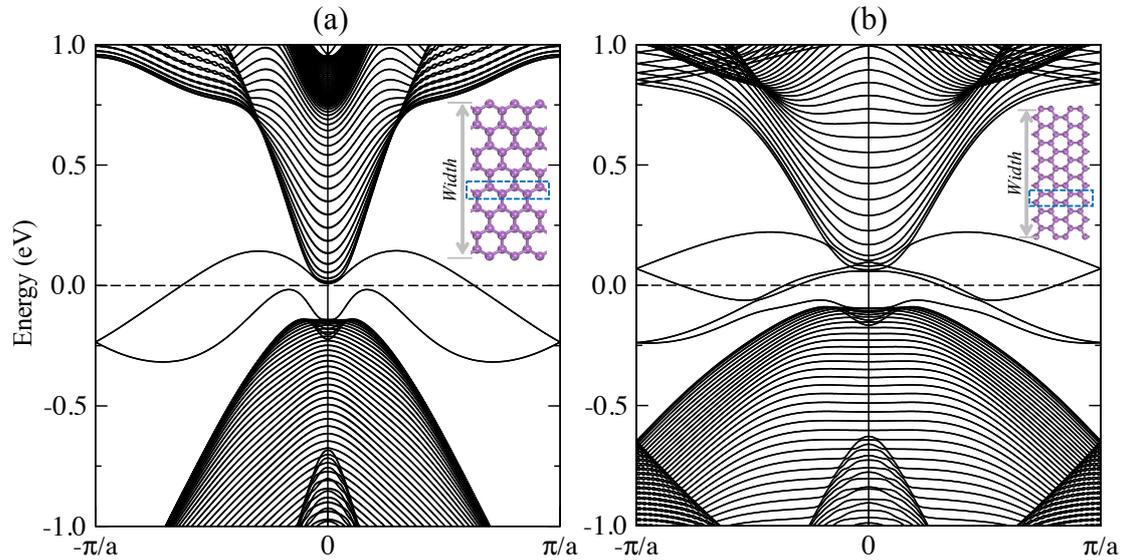

Fig. 2. Band structures of 1D Bi nanoribbons with (a) zigzag and (b) armchair edges. The Fermi level (dashed horizontal line) is set as the zero energy. The insets present the atomic structures of Bi nanoribbons with a width of 8 Bi chains. The blue dashed rectangles in (a) and (b) mark one zigzag and armchair Bi chain, respectively. The actual width of the nanoribbons for the band structures is 60 Bi chains.

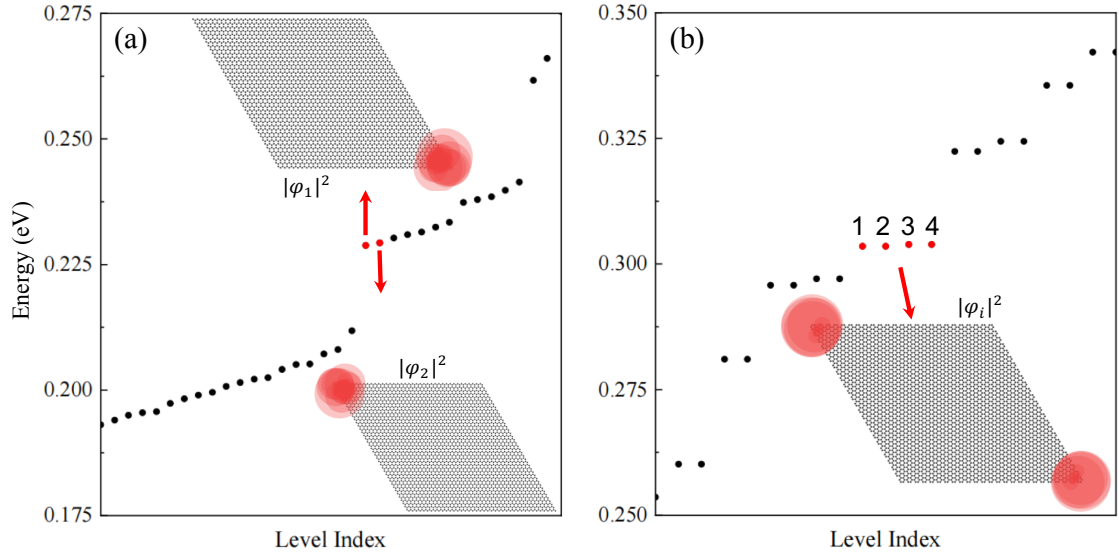

Fig. 3. Energy levels of Bi rhombic nanoflakes with (a) zigzag and (b) armchair edges with respect to the Fermi level. The insets display the spatial distributions of the probability densities of the corner-state wavefunctions ($|\varphi_i|^2$). The widths of the zigzag and armchair nanoflakes are 40 and 25 Bi chains, respectively.

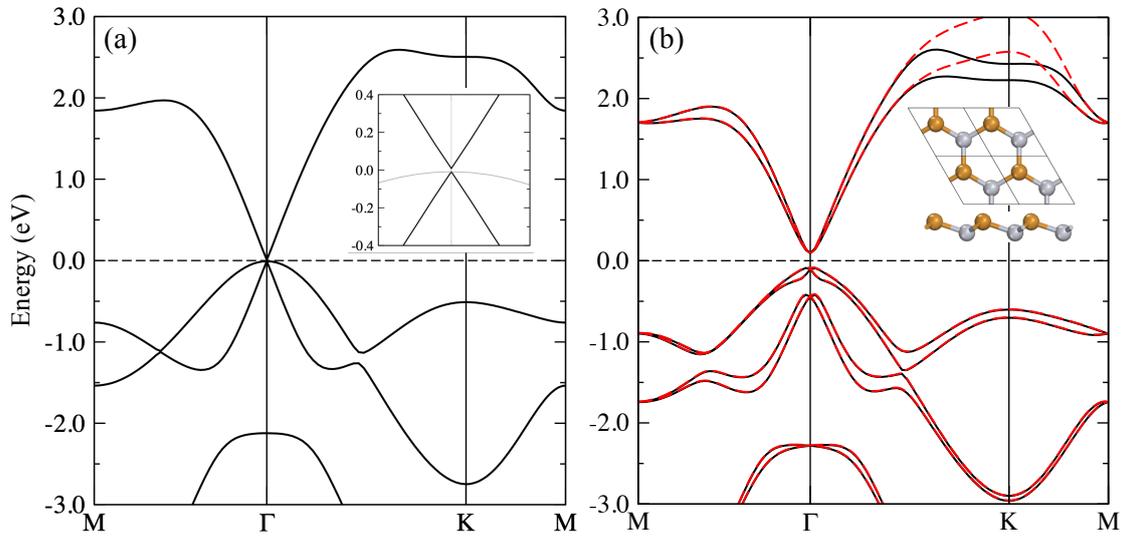

Fig. 4. Band structures of 2D honeycomb HgTe (a) without and (b) with spin-orbit coupling. The red dashed curves in (b) are the energy bands from tight-binding model with the fitted basis set of Wannier functions. The inset in (b) zooms in the energy bands near the Γ point. The inset in (b) presents the atomic structure of 2D honeycomb HgTe, where the grey and brown spheres stand for Hg and Te atoms, respectively.

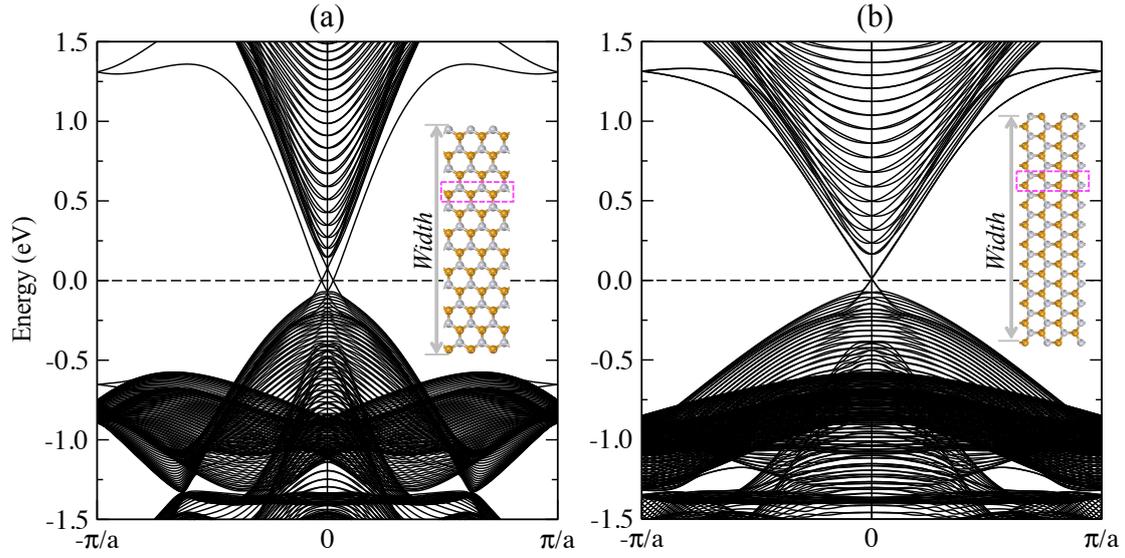

Fig. 5. Band structures of 1D HgTe nanoribbons with (a) zigzag and (b) armchair edges. The insets present the atomic structures of HgTe nanoribbons with a width of 12 Hg-Te chains. The purple dashed rectangles in (a) and (b) mark one zigzag and armchair Hg-Te chain, respectively. The actual width of the nanoribbons for the band structures is 60 Hg-Te chains.

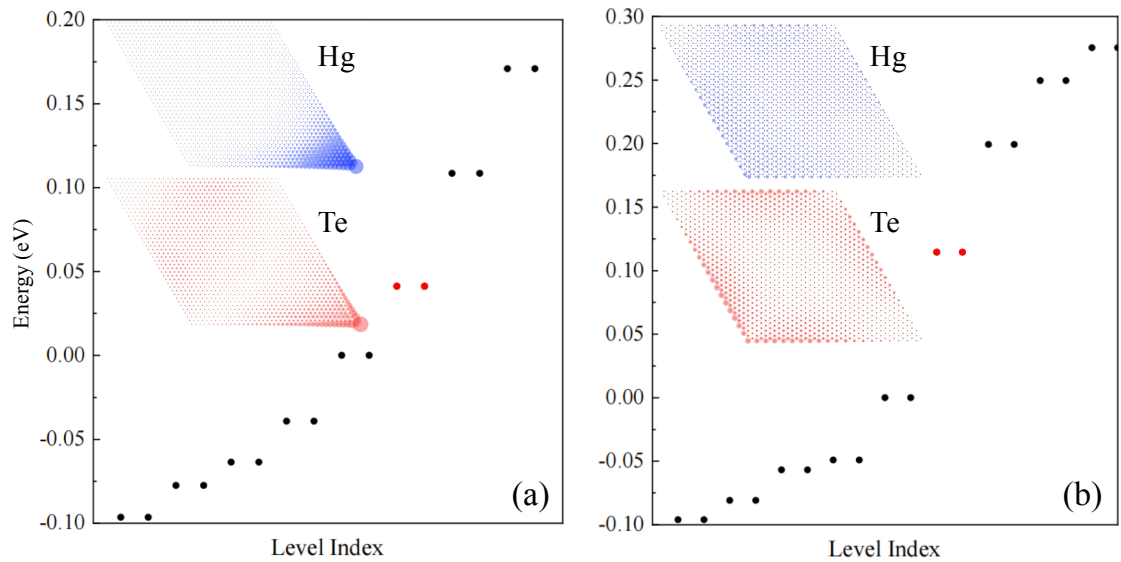

Fig. 6. Energy levels of HgTe rhombic nanoflakes with (a) zigzag and (b) armchair edges, with respect to the Fermi level. The insets display the spatial distributions of the probability densities of the corner-state wavefunctions ($|\varphi_i|^2$) projected on Hg and Te atoms. The widths of the zigzag and armchair nanoflakes are 40 and 25 Hg-Te chains, respectively.

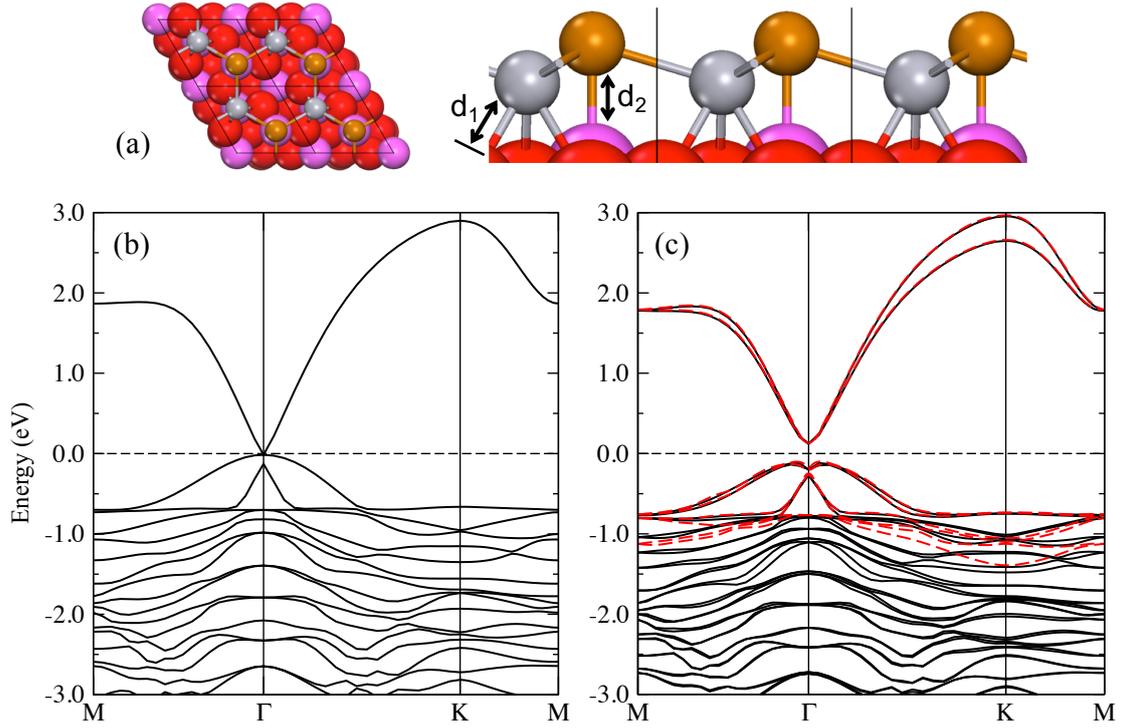

Fig. 7. Structural and electronic properties of 2D honeycomb HgTe on $Al_2O_3(0001)$ substrate. (a) Top and side views of the atomic structure. The grey, brown, purple and red spheres stand for Hg, Te, Al and O atoms, respectively. The rhombuses indicate the unit cells. (b) and (c) Band structures without and with spin-orbit coupling, respectively. The red dashed curves in (c) are the energy bands from tight-binding model with the fitted basis set of Wannier functions.

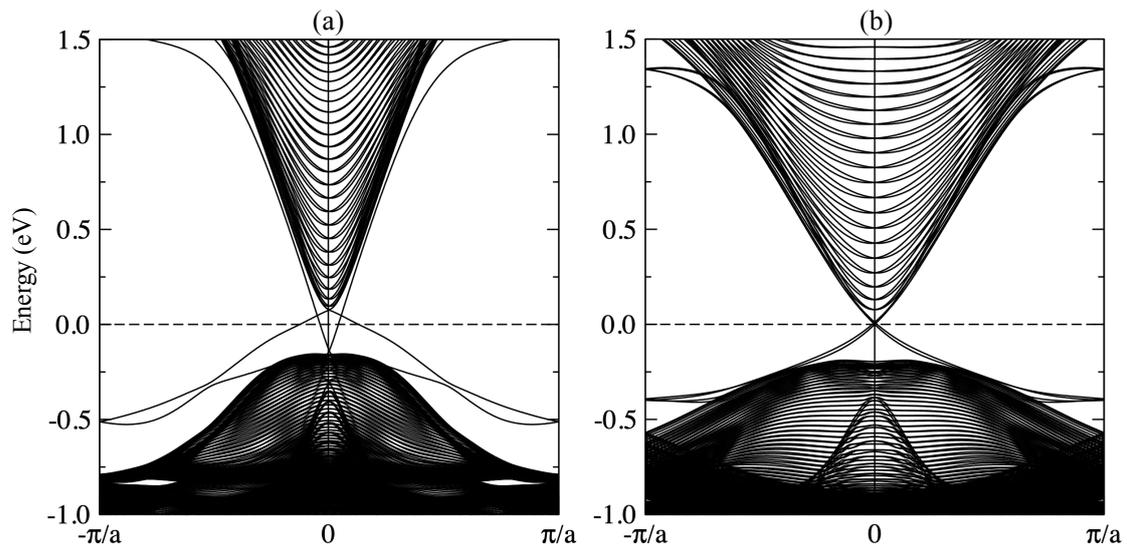

Fig. 8. Band structures of 1D HgTe nanoribbons on $Al_2O_3(0001)$ substrate with (a) zigzag and (b) armchair edges. The widths of the nanoribbons is 60 Hg-Te chains, the same as those in freestanding nanoribbons in Fig. 5.

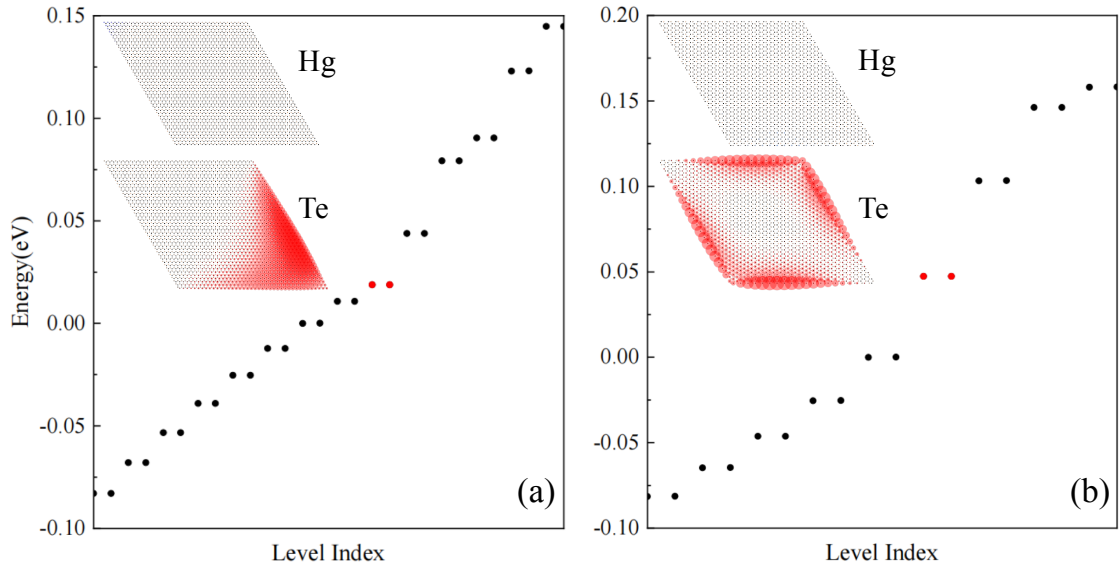

Fig. 9. Energy levels of HgTe rhombic nanoflakes on $Al_2O_3(0001)$ substrate with (a) zigzag and (b) armchair edges, with respect to the Fermi level. The insets display the spatial distributions of the probability densities of the corner-state wavefunctions ($|\varphi_i|^2$) projected on Hg and Te atoms. The widths of the zigzag and armchair nanoflakes are 40 and 25 Hg-Te chains, respectively.